\documentclass[12pt]{article}
\usepackage{amsmath,epsfig,cite,pstricks}

\newcommand{\tev}{\,\, \mathrm{TeV}}
\newcommand{\gev}{\,\, \mathrm{GeV}}

\newcommand{\RR}{{\rm R}}
\newcommand{\LL}{{\rm L}}
\newcommand{\tr}{{\rm tr}}
\newcommand{\gesim}{\,{_{\textstyle >}\atop^{\textstyle\sim}}\,}

\begin{document}
\thispagestyle{empty}

\begin{center}

{\large\bf  Phenomenology of the Little Higgs model with X-Parity}

\vspace*{3ex}

{\sc
A.~Freitas$^1$, P.~Schwaller$^2$,
D.~Wyler$^2$
}

\vspace*{2ex}

{\sl\small $^1$
Department of Physics \& Astronomy, University of Pittsburgh,\\
3941 O'Hara St, Pittsburgh, PA 15260, USA
}
\\[1ex]
{\sl\small $^2$
Institut f\"ur Theoretische Physik,
        Universit\"at Z\"urich, \\ Winterthurerstrasse 190, CH-8057
        Z\"urich, Switzerland
}
       
\end{center}

\begin{abstract}
In the popular littlest Higgs model, T-parity can be broken by
Wess-Zumino-Witten (WZW) terms induced by a strongly coupled UV completion. On
the other hand, certain models with multiple scalar multiplets (called moose
models) permit the implementation of an exchange symmetry (X-parity) such that it is not broken by the
WZW terms. Here we present a concrete and realistic construction of such a
model. The little Higgs model with X-Parity is a concrete and realistic implementation of this idea.
 In this contribution, the properties of the model are reviewed and the collider phenomenology is discussed in some detail.
We also present new results 
on the decay properties and LHC signatures of the light triplet scalars that are predicted by this model. 
\end{abstract}

% typeset front matter (including abstract)

\subsubsection*{1\hspace{1em} THE LITTLE HIGGS AND T-PARITY}

Little Higgs models \cite{lh} can naturally explain a relatively large hierarchy
between the electroweak scale and a fundamental symmetry breaking scale that is
several orders of magnitude larger, $\Lambda \sim 10$~TeV. In these models, the
Higgs boson is a Goldstone boson of a spontaneously broken global symmetry. In
order to construct realistic models with a non-trivial Higgs potential and
fermion Yukawa couplings, the global symmetry needs to be broken explicitly, and
thus can only be an approximate symmetry. Nevertheless, if each of these
explicit breaking terms keeps a subgroup of the global symmetry unbroken, a
Higgs mass term is only generated beyond one-loop order and thus can naturally
be small.

Little Higgs models predict additional particles with masses at an intermediate
scale $f \sim 1$~TeV. These new particles lead to large tree-level corrections
to electroweak precision observables \cite{ewpo}, which rule out values of $f$
below several TeV. However, the introduction of a discrete symmetry called
T-parity \cite{tpar,LHT} forbids the dangerous tree-level diagrams. As a result,
models with T-parity are in agreement with precision data for values of $f$ even
below 1~TeV \cite{Hubisz:2005tx}. Moreover, the lightest T-odd particle would be
stable and thus could be a candidate for dark matter.

In Ref.~\cite{hillsq} it was argued that this picture is not complete, but that
a Wess-Zumino-Witten (WZW) term \cite{WZW}
should be added to the Lagrangian of the model.
WZW terms typically occur if the symmetry breaking at the scale $\Lambda$ is
facilitated by some strong
dynamics. As was shown in Ref.~\cite{hillsq}, this WZW term would be odd under
T-parity, leading to the decay of the lightest T-odd particle \cite{tdec}.
However, recently it was shown that a parity can be implemented in a different
way such that that it leaves the WZW term invariant \cite{ky}. In this approach
the discrete symmetry, which will be named X-parity henceforth, is defined as a
exchange symmetry between two scalar fields. This idea has been worked out into
a complete and realistic model \cite{mmx}, called the minimal moose model
with X-parity. In this contribution, the properties of the model will be reviewed and the collider phenomenology will be discussed in some detail. We also present new results 
on the decay properties and LHC signatures of the light triplet scalars that are predicted by this model.

\subsubsection*{2\hspace{1em} THE MODEL}

\begin{figure}[h]
\begin{tabular}{cccc}
\hspace{2cm} &
\parbox[b]{5cm}{%
\psfig{figure=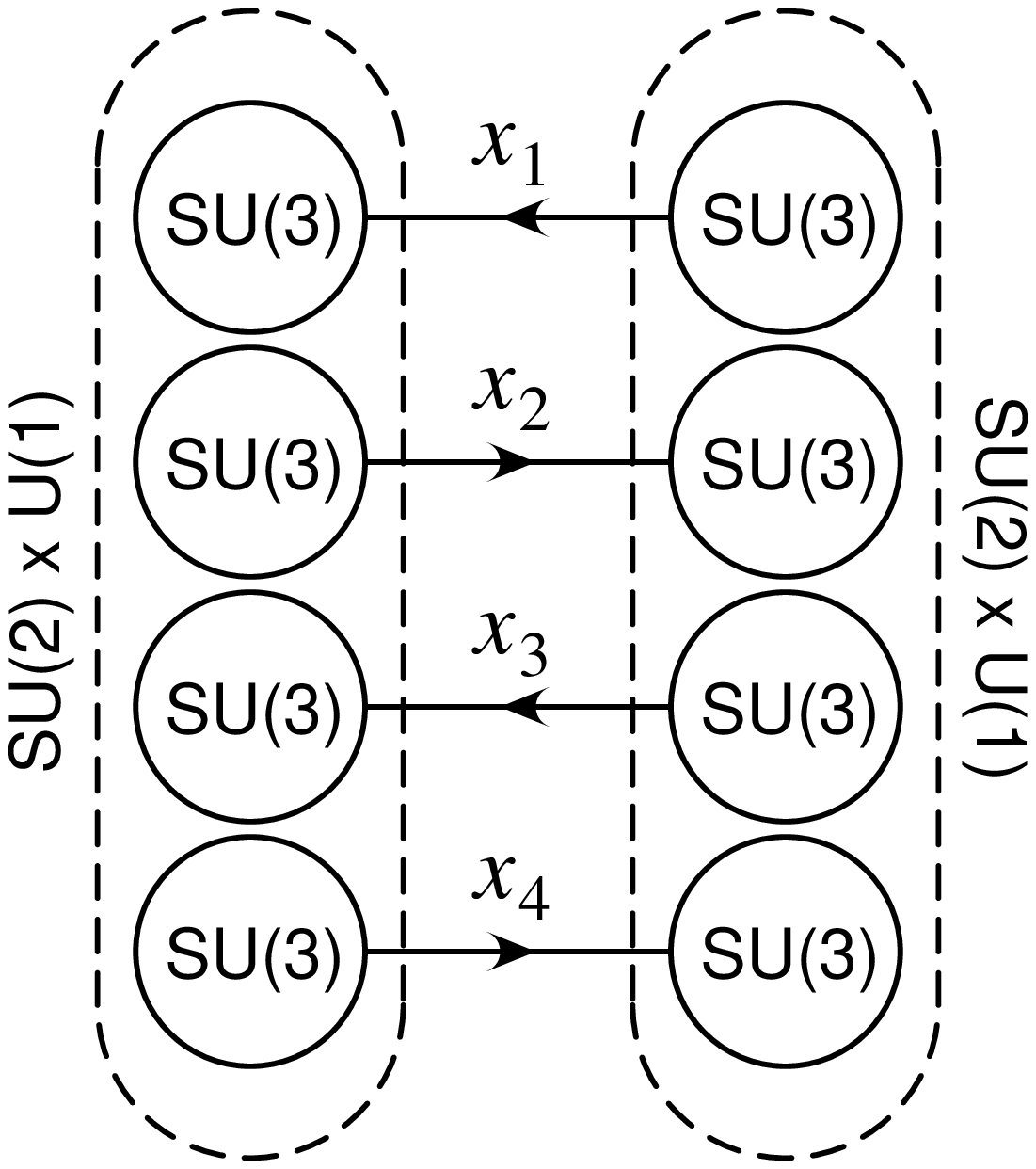, width=5cm}\\[1.7em]
\psfig{figure=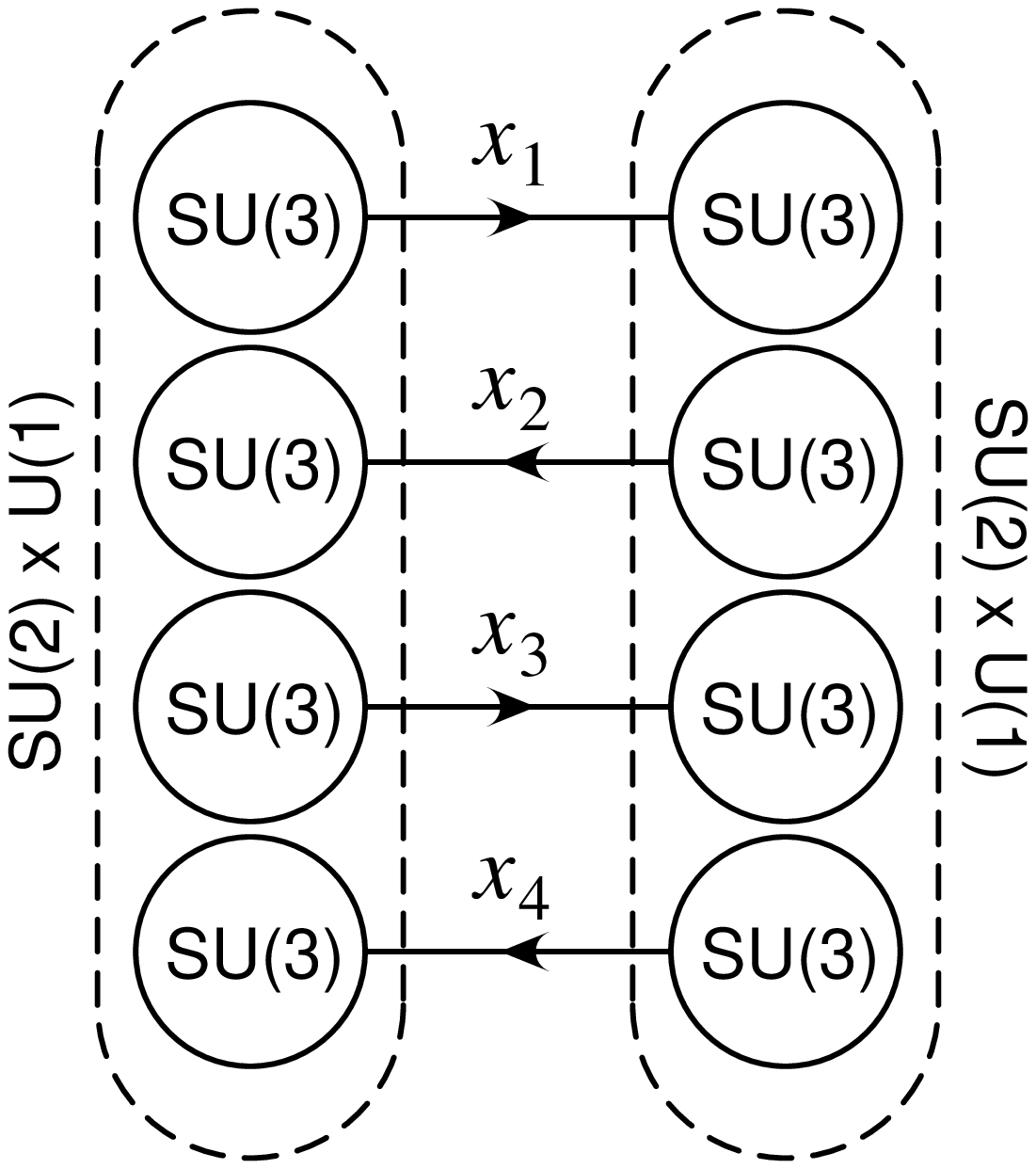, width=5cm}%
\psline[linewidth=2pt,arrowsize=4pt 3,linecolor=red]{->}(-3.8,5.2)(-2.1,4.4)%
\psline[linewidth=2pt,arrowsize=4pt 3,linecolor=red]{->}(-2.1,5.2)(-3.8,4.4)%
}
& \hspace{.5cm} &
\parbox[b]{7cm}{%
\caption{Illustration of the global and gauge symmetry structure of the model.
The difference between the upper and the lower diagram and the red arrows indicate
the effect of X-parity.}\label{moosediag}}
\end{tabular}
\end{figure}

The model is
constructed as an effective theory valid up to the cutoff scale $\Lambda \sim
10$~TeV. At this scale some unspecified new physics is assumed to be
responsible for the fundamental symmetry breaking and for a renormalizable
completion of the theory.
The model is based on
a [SU(3)$_{\rm L}\times$SU(3)$_{\rm R}$]$^4$ global symmetry, which is spontaneously broken to the
diagonal subgroup SU(3)$^4$ by the fundamental symmetry-breaking mechanism. As a result, four copies of a non-linear sigma
model appear, $X_i = e^{2i x_i/f}$, $i=1,2,3,4$. Under the global symmetry they
transform as
\begin{equation}
X_{1,3} \to L_{1,3} X_{1,3} R^\dagger_{1,3},
\;\;\;
X_{2,4} \to R_{2,4} X_{2,4} L^\dagger_{2,4}, 
\end{equation}
which can be interpreted as the $X_{1,3}$ and $X_{2,4}$ having ``opposite
directions'' in a moose diagram. The scalar sector, defined in this way, is
invariant under the exchange symmetry
\begin{equation}
{\rm SU(3)}_{\rm L} \leftrightarrow {\rm SU(3)}_{\rm R},
\quad
X_1 \leftrightarrow X_2,
\quad
X_3 \leftrightarrow X_4.
\end{equation}
This symmetry will be called X-parity in the following. It has the remarkable
feature that is leaves the WZW term invariant \cite{ky}, so that the X-parity is
an exact symmetry and the lightest X-odd particle is stable. 
The symmetry structure and the effect of X-parity is depicted in terms of a
moose diagram in Fig.~\ref{moosediag}.

In order to construct a complete model, a $[SU(2) \times U(1)]_{\rm L} \times
[SU(2) \times U(1)]_{\rm R}$ subgroup is gauged, resulting in a set of heavy X-odd gauge
bosons with masses of order $f \sim 1$~TeV and a set of massless X-even gauge
bosons that are identified with the Standard Model (SM) gauge bosons. 
The scalar
Lagrangian is given by
\begin{align}
\label{LG}
{\cal L}_{\rm G} &= \frac{f^2}{4} \sum_{i=1}^4 \tr [(D_\mu X_i)(D^\mu
X_i)^\dagger], & \text{with}&&
&D_{\mu} X_{1,3} = \partial_{\mu}X_{1,3} - i A_{\LL\mu} X_{1,3}
                                        + i X_{1,3}A_{\RR\mu}, \nonumber
\\[-2ex]
&&&& \nonumber
&D_{\mu} X_{2,4} = \partial_{\mu}X_{2,4} - i A_{\RR\mu} X_{2,4}
                                        + i X_{2,4}A_{\LL\mu}.
\end{align}
Under X-parity the gauge fields transform as $A_L \leftrightarrow A_R$.

To implement X-parity in the fermion sector, we introduce two left-handed
fermion doublets for each fermion flavor, illustrated here for the quark sector.
They are embedded into incomplete representations of SU(3) as follows
\begin{equation}
  Q_{\rm a}=(d_{\rm a}, u_{\rm a}, 0)^\top, \qquad\qquad  
  Q_{\rm b}=(d_{\rm b}, u_{\rm b}, 0)^\top.
  \nonumber
\end{equation}
Under the global $SU(3)_\LL\times SU(3)_\RR$ group they transform as $Q_{\rm a}
\to L_i Q_{\rm a}$ and $Q_{\rm b} \to R_i Q_{\rm b}$, while X-parity
interchanges the two fields, $Q_{\rm a} \leftrightarrow Q_{\rm b}$. The kinetic
terms for these fermions assume the standard form. Under X-parity, they
decompose into the X-even standard model fermions and their X-odd partners, the
mirror fermions, denoted by $Q_H$. They can be given an ${\cal O}(f)$ Dirac mass
by the following parity invariant mass term:
\begin{equation}
  {\cal L}_{\rm M} = -\frac{\lambda_{c}}{\sqrt{2}} f \left( Q_{\rm a} \xi_{1} Q_{\rm c}^{c} 
        - Q_{\rm b}\Omega \xi_{1}^{\dagger} Q_{\rm c}^{c} 
        - Q_{\rm b}\xi_{2}\Omega Q_{\rm c}^{c} 
        + Q_{\rm a}\Omega \xi_{2}^{\dagger}\Omega Q_{\rm c}^{c}\right) 
        + \mathrm{h.c.}\, \nonumber
\end{equation}
This term introduces the righthanded partners $Q_c^c=(d_c^c,u_c^c,0)^T$ for the
X-odd fermions $Q_H$ and $\xi_i = \sqrt{X_i} = e^{i x_i/f}$. 

The Yukawa couplings of the light quarks and leptons are small and therefore
don't lead to large contributions to the Higgs mass through radiative
corrections. This is however not true for the top quark and therefore the top
quark Yukawa coupling requires special attention. To not explicitly break the
global symmetries, the third family doublets have to be embedded into complete
$SU(3)$ triplets by adding a third vector-like quark:
\begin{equation}
  Q_{\rm 3a}=(d_{\rm 3a}, u_{\rm 3a}, U_{\rm a})^\top, \qquad
  Q_{\rm 3b}=(d_{\rm 3b}, u_{\rm 3b}, U_{\rm b})^\top, \qquad 
  Q_{\rm 3c}^c=(d_{\rm 3c}^c, u_{\rm 3c}^c, U_{\rm c}^c)^\top. \nonumber
\end{equation}
The top Yukawa coupling now is given by 
\begin{equation}
{\cal L}_{\rm t} = -\lambda f 
  Q_{\rm 3a} (X_3 + \Omega X_4^\dagger\Omega) 
        \begin{pmatrix} 0 \\ 0 \\ U_{\rm b}^c \end{pmatrix}
- \lambda f Q_{\rm 3b} (\Omega X_3^\dagger\Omega + X_4) 
        \begin{pmatrix} 0 \\ 0 \\ U_{\rm a}^c \end{pmatrix} + \mathrm{h.c.} \,,
        \nonumber
\end{equation}
where we have introduced two additional singlets $U^c_a$, $U^c_b$ that transform
as $U^c_a \leftrightarrow U^c_b$ under X-parity, and the X-even combination of
which will form the right-handed component of the top quark. 

In
total the top quark has three partners, two of which are X-odd, and one being
X-even. The ratio $R=\lambda/\lambda_c$ of the two top Yukawa couplings is a
free parameter of the model. 
% 
% We have checked \cite{mmx} that the model can lead to successful
% electroweak symmetry breaking for
% natural parameter choices, {\it i.$\,$e.} without fine-tuning beyond the
% few-percent level.

\subsubsection*{3\hspace{1em} MASS SPECTRUM}

All X-odd partners
of the standard model fermions have ${\cal O}(f)$ masses whose precise values
are determined by the parameter $\lambda_c$. An exception is the top-quark
sector, where we have in total four states, two X-even and two X-odd fermions.
While both X-odd fields obtain ${\cal O}(f)$ masses, the X-even fields mix to
give the physical top-quark that remains massless before EWSB and an additional
$T$-quark with a mass of ${\cal O}(f)$. The mixing in the top sector is
determined by a single parameter $R=\lambda/\lambda_c$ that is the ratio of the
top Yukawa coupling and the mirror fermion mass parameter.

The masses of the X-odd gauge bosons are fixed once the symmetry breaking scale
$f$ is specified. If $\lambda_c \sim {\cal O}(1)$, the X-odd partner of the
hypercharge boson, $A_H$, is always the lightest heavy gauge boson, which can
serve as a dark matter candidate.

The scalar pseudo-Goldstone fields can be decomposed into SU(2) gauge multiplets
as follows: \begin{equation} x_i = \begin{pmatrix} \phi_{i}+ \eta_{i}/\sqrt{12}
& h_{i}/2 \\ h_{i}^{\dagger}/2 & - \eta_{i} / \sqrt{3} \end{pmatrix},
\end{equation} where $\phi_i=\phi_i^a\sigma^a/2$ are SU(2) triplets, $h_i$ are
complex doublets, and $\eta_i$ are real singlets. The explicit partial breaking
of the global symmetry through the Higgs self-interaction, the gauge
interactions, the mirror fermion kinetic and mass terms, and the top Yukawa
coupling generate masses for all of these scalar particles at the one- and
two-loop level. These loop contributions are divergent and depend on the cut-off
scale $\Lambda$ of the effective theory. Nevertheless, the order of magnitude of
the mass parameters can be estimated parametrically, leading to ${\cal O}(f)
\sim {\cal O}$(1~TeV) masses for almost all scalars. Only one X-even triplet,
called $\phi_a$, and
one X-even doublet obtain masses of the order of the electroweak scale, ${\cal
O}$(100~GeV). 

The light doublet will form the SM-like ``little'' Higgs boson. Together with
the second, heavy X-even doublet, the structure of the Higgs potential is
similar to a Two-Higgs-Doublet model. However, mixing between the two X-even
doublets is suppressed by their mass ratio ${\cal O}$(100~GeV)$^2/{\cal
O}$(1~TeV)$^2$. Nevertheless, mixing effects could lead to interesting consequences,
since they have been shown to necessarily lead to CP violation \cite{mmx}.

Table~\ref{part} summarizes the particle content beyond the SM fermions and gauge
bosons. Also indicated in the table are the quantum numbers of these fields with
respect to the X-parity. Note that the present model is very similar to the
older moose model proposed in Ref.~\cite{LHT}. As a result, the original
T-parity proposed by Cheng and Low \cite{LHT} is also respected by our model on
the classical level, although it is broken by the WZW term as explained above.
However, since the WZW term is the only source of T-parity breaking, T-parity is
still a good quantum number for most production and decay processes, and
therefore it is also listed in Tab.~\ref{part}.

\begin{table}[tb]
\caption{Particle content of the minimal moose model with X-parity (besides SM
fermions and gauge bosons). Also listed is the even-/oddness of the particles
under X- and T-parity.}
\label{part}
\renewcommand{\arraystretch}{1.2} % enlarge line spacing
\begin{tabular}[t]{lc@{\hspace{2em}}cc}
\hline
    Field & & X-parity & T-parity\\
\hline\hline
Heavy gauge bosons & $B_{H}^0,W_{H}^{0,\pm}$ & $-$ & $-$\\
\hline
Singlet scalars & $\eta_x$ & $-$ & $-$ \\
 & $\eta_a,\eta_b$ & $+$ & $-$ \\
\hline
Triplet scalars & $\phi_x$ & $-$ & $-$ \\
 & $\phi_a,\phi_b$ & $+$ & $-$ \\
\hline
X-odd doublets & $h_{H1},h_{H2}$ & $-$ & $+$ \\
\hline
X-even doublets  & \hspace{-1.5em}$h^0,H^0, A^0, H^\pm$ & $+$ & $+$ \\
\hline
Mirror leptons & $L_H$ & $-$ & $-$ \\
\hline
Mirror quarks & $Q_H$ & $-$ & $-$ \\
\hline
Top partners & $T_H,T'$ & $-$ & $-$ \\
 & $T$ & $+$ & $+$ \\
\hline
\end{tabular}
\end{table}

\begin{figure*}[tb]
\fbox{\psfig{figure=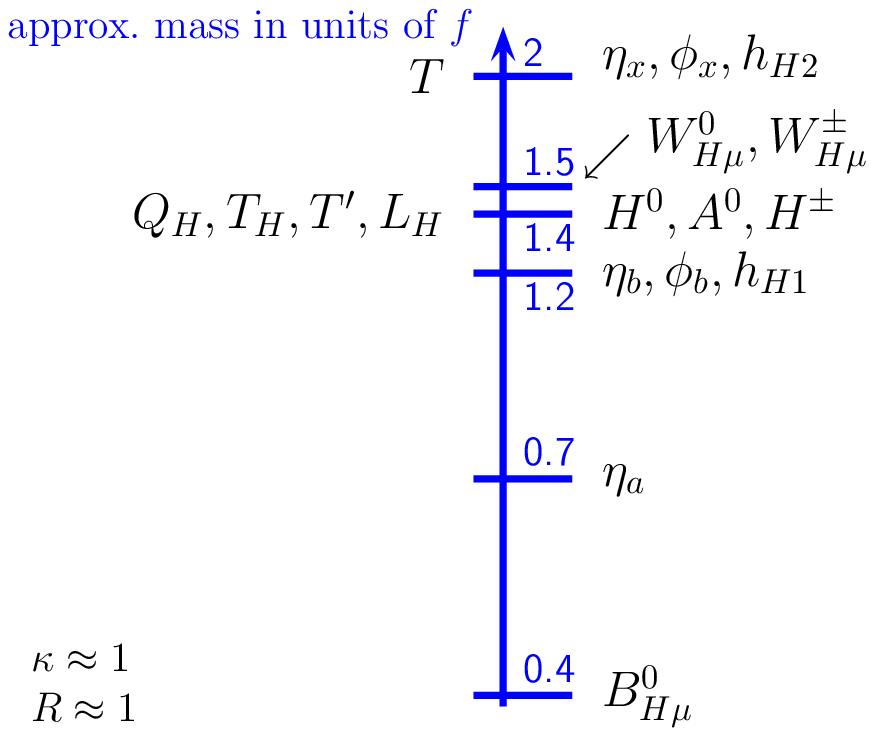, width=3in, bb=170 465 460 680}}
\fbox{\psfig{figure=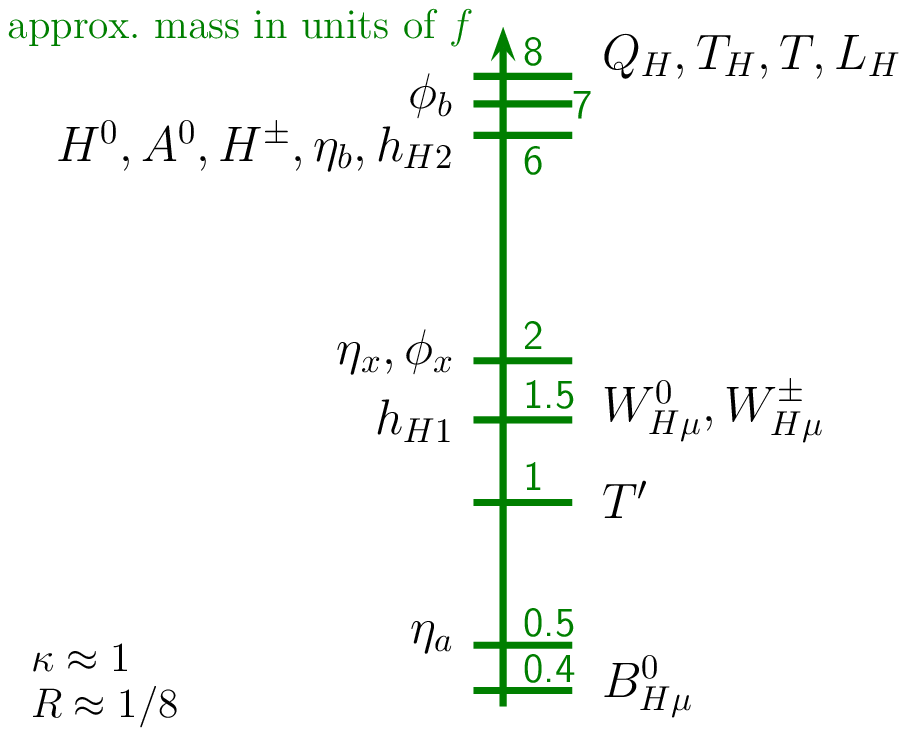, width=3in, bb=170 465 460 680}}
%\vspace{-1ex}
\caption{Approximate patterns of two typical mass spectra for heavy new
particles.}
\label{masspat}
\end{figure*}

As mentioned above, the particle spectrum cannot be predicted precisely due to
the dependence on the unknown UV physics at the scale $\Lambda$. Within the
framework of the effective theory, the UV completion will generate counterterms
that cancel the divergence of the loop corrections to the mass parameters.
Setting all unknown coefficients of the finite parts of these counterterms to
one, the estimates for the mass patterns in Fig.~\ref{masspat} are
obtained, for two choices of the Yukawa ratio $R$.

For ${\cal O}(1)$ values of the unknown counterterm coefficients, the lightest
X-odd particle is the heavy U(1) gauge boson $B_{H}^0$. Since it is weakly
interacting and stable it is a viable dark matter candidate.

Note that the SM-like Higgs boson $h^0$ and the X-even scalar triplet $\phi_a$
are not shown in Fig.~\ref{masspat}, since they have have masses of the order of
the electroweak scale, as mentioned above, and thus are much lighter than the
other new particles.

\subsubsection*{4\hspace{1em} ELECTROWEAK PRECISION CONSTRAINTS}

To estimate the effect of the new particles on the electroweak precision
constraints, we calculate the leading contributions to the T-parameter. An
important positive contribution to T comes from the custodial symmetry violating
$|h^\dagger D_\mu h|^2$ operators that are contained in the kinetic term for the
Goldstone fields. This is partially cancelled by a negative contribution from
the mass splitting between the X-odd gauge bosons $W^\pm_H$ and $W^0_H$ and by a
contribution from the Higgs sector of the model that can be either positive or
negative, both appearing at the one-loop level. The mixing in the top sector
leads to a contribution of similar size at one loop that strongly depends on the
parameter $R$. 

\begin{figure}[tb]
%\begin{minipage}[b]{7.9cm}
\includegraphics[width=.47\textwidth]{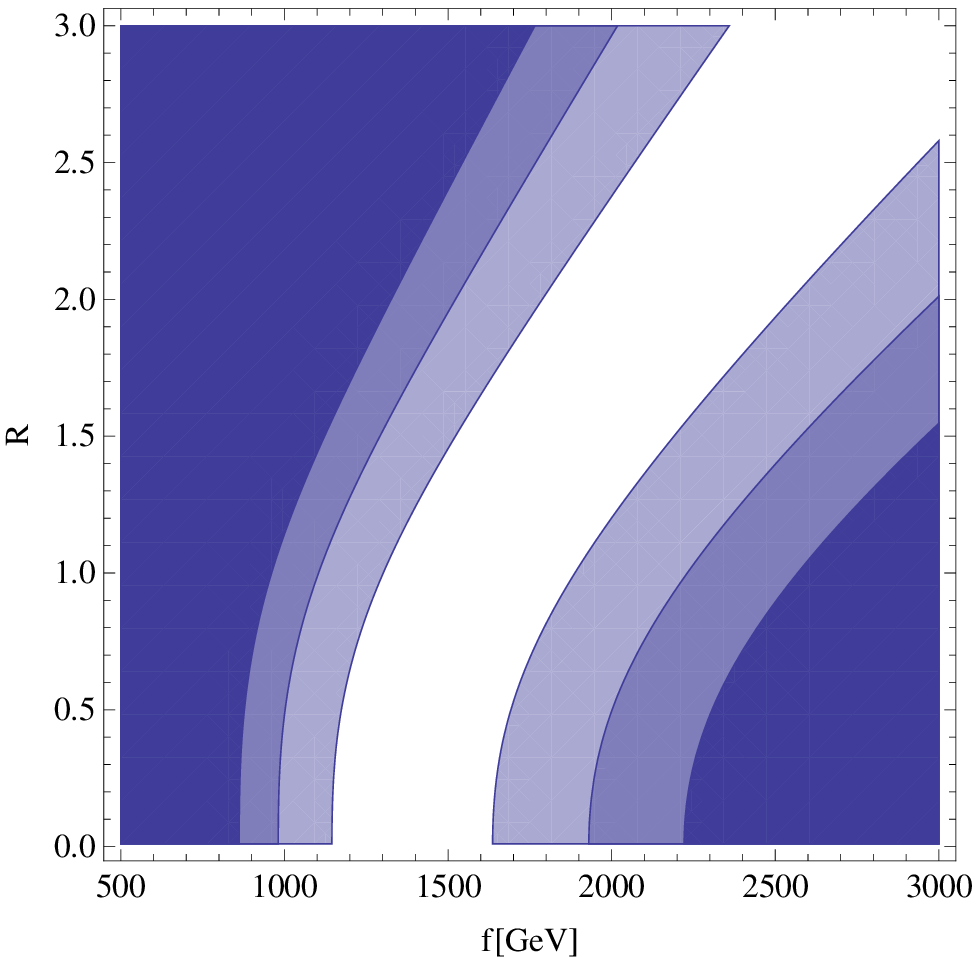} \hfill
%\end{minipage}%
%\hfill%
%\begin{minipage}[b]{8cm}
\includegraphics[width=.47\textwidth]{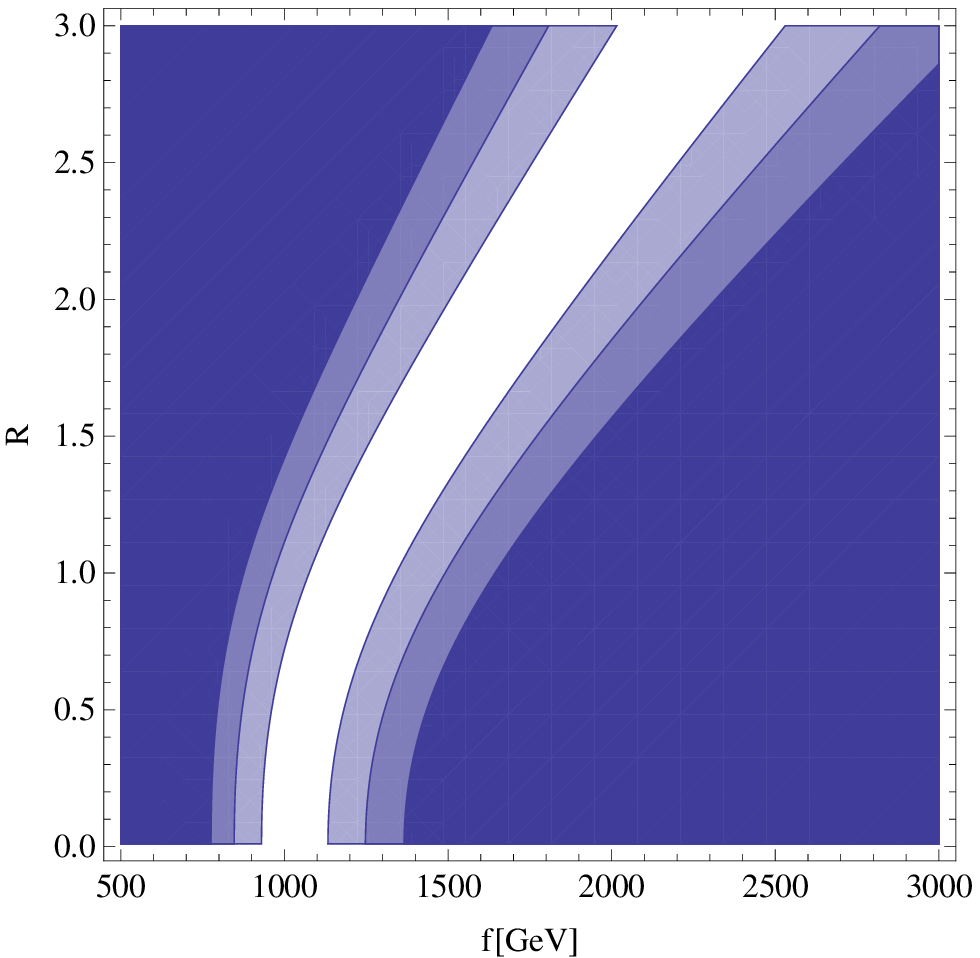}
%\end{minipage}
\vspace{-.5em}
\caption{Allowed regions in the $f-R$ plane where the T-parameter agrees with
the experimental value. In the left plot we use  $\delta_\pm^2 = 0.1 f^2$ and
$\delta_0^2 = 0.2 f^2$ while the right plot uses $\delta_\pm^2 = -0.15 f^2$ and
$\delta_0^2 = -0.3 f^2$, see \cite{mmx} for details.  
\label{fig1}}
%\vspace*{-\baselineskip}
\end{figure}

All contributions to T are suppressed by the scale $f$ and essentially decouple
for $f\geq 2\tev$. Lower values of $f$ are viable provided some moderate
cancellations happen. Figure \ref{fig1} shows the allowed region in the $f-R$
plane with all other parameters fixed. We see that values of $f$ down to the TeV
scale and slightly below are possible, making the model testable at the LHC.

\subsubsection*{5\hspace{1em} COLLIDER PHENOMENOLOGY}

Due to the exact realization of X-parity, X-odd particles can be produced only
in pairs at colliders. They then decay in cascades that eventually end with the
lightest X-odd particle, the $B_{H}^0$ boson. Assuming $f \sim 1$~TeV, most of
the new particles are within reach of the Large Hadron Collider (LHC), but
separation of the small signals from the background is difficult in many cases.

For instance, the heavy SU(2) gauge bosons $W_{H}^{0,\pm}$ are predicted to be
much heavier than the lighter $B_{H}^0$ boson, and since they only have
electroweak couplings, their production cross sections are relatively small
\cite{lhtpheno}. The same is true for most of the heavy scalar particles with
${\cal O}(f) \sim {\cal O}$(TeV) masses.

\begin{figure}[htb]
\centering
\epsfig{figure=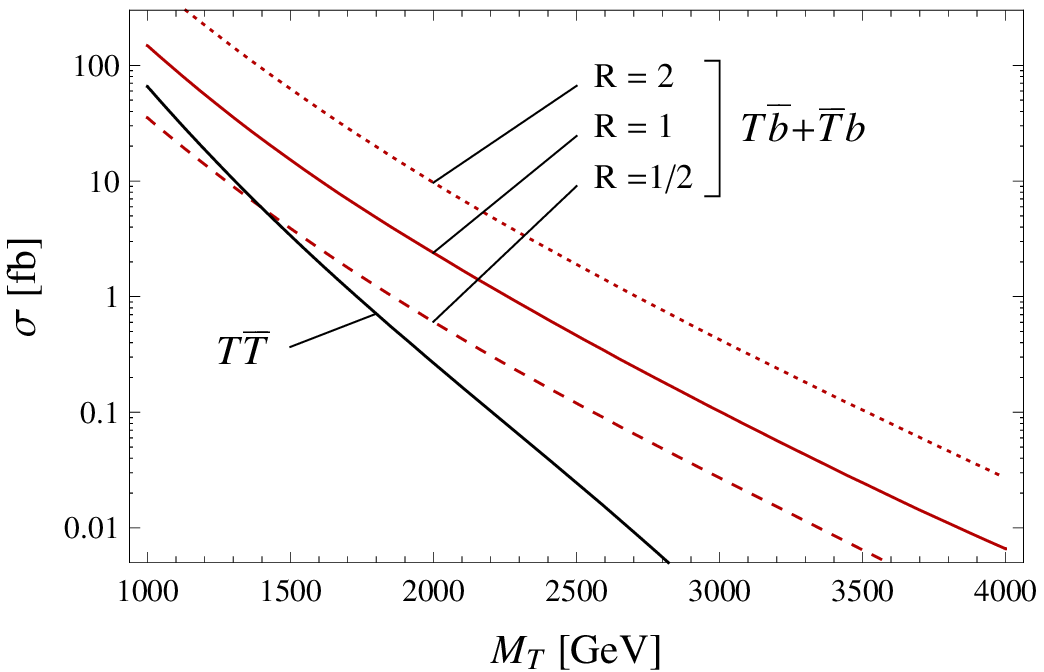, width=10cm}%
\vspace{-.3em}
\caption{LHC cross sections for single $T$ and $T\overline{T}$ production, as a
function of the $T$ quark mass, and for different values of
$R\equiv\lambda_1/\lambda_2$. The QCD and factorization scales have been
set to $M_{T}$, and the center-of-mass energy is
$\sqrt{s} =14$~TeV.}
\label{T}
\end{figure}

Relatively large cross sections are expected for the colored particles. In
particular, the X-even top partner $T$ can be produced singly.
Single $T$ production, $pp \to T\bar{b}+X, \; \bar{T}b+X$ proceeds dominantly
through the partonic processes $b\bar{q} \to T \bar{q}'$ and $\bar{b}q \to
\overline{T} q'$, where $q,q'$ are SM quarks of the first two generations. The 
requirement of bottom quarks in the initial state, originating from the parton
distribution function of the proton, leads to a suppression of the single
production process, so that pair production through $gg \to T\overline{T}$ and
$q\bar{q}\to T\overline{T}$ can be competitive. The LHC production cross sections are shown in
Fig.~\ref{T}.

The single $T$ production cross section depends sensitively on the mixing
between the ordinary top quark $t$ and the heavy $T$ quark, and thus on the
parameter $R\equiv\lambda_1/\lambda_2$. On the other hand, the pair production
process is mainly mediated through QCD interactions and therefore insensitive to
the mixing parameters. In spite of the suppression, the single production
process is dominant for $M_T \gesim 1$~TeV.

\begin{figure}[htb]
\centering
\epsfig{figure=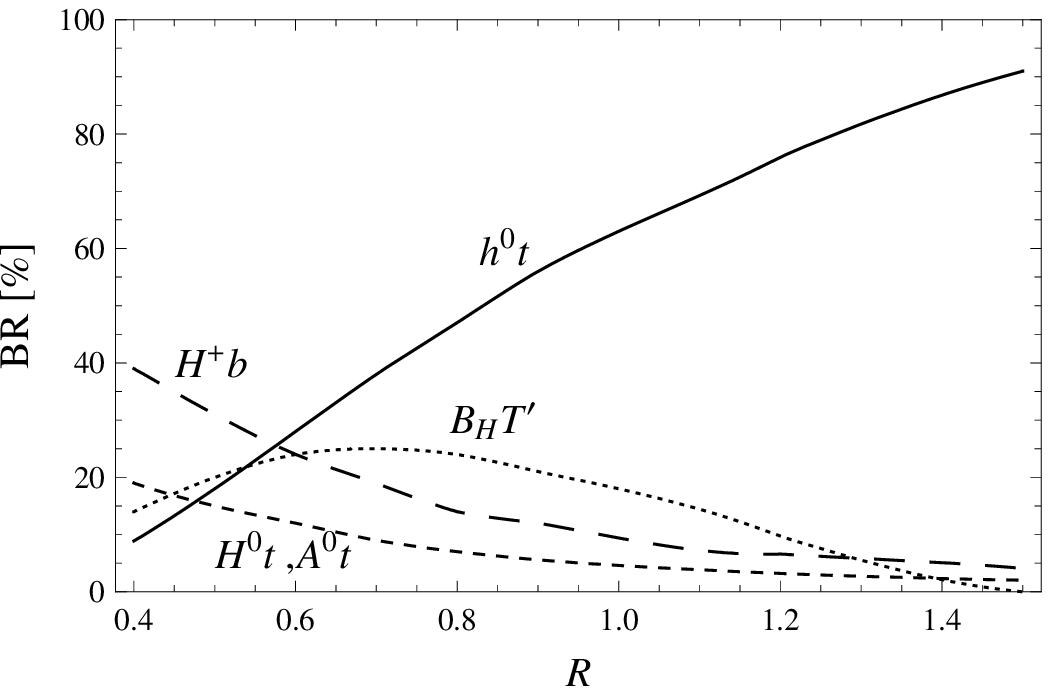, width=8.9cm}%
\vspace{-1em}
\caption{Branching ratio of the heavy X-even $T$ quark, as a function of the
ratio $R=\lambda_1/\lambda_2$ of top Yukawa couplings.}
\label{BRT}
\end{figure}

The $T$ quark can decay into Higgs bosons via the top Yukawa couplings or into
the $B_H^0$ boson via its gauge coupling. The branching ratios are shown in
Fig.~\ref{BRT}. For most of the parameter space the channel $T\to h^0 t$ is
dominant. If the $h^0$ Higgs boson is light, and thus mainly decays via $h^0\to
b\bar{b}$, the single $T$ mode leads to a final state signature of $4b+W$. The
separation of this signal from the SM background will rely heavily on b-tagging
and requires a dedicated analysis.

However, the X-odd mirror quarks will be even more difficult to detect at the
LHC. They are produced only in pairs, so that the cross sections are smaller
than for the $T$ quark. Furthermore, due to the large mass of the SU(2) gauge
bosons $W_H^{0,\pm}$, mirror quarks decay directly to the lightest X-odd
particle via $Q_H \to q \, B_H^0$ for $R \sim {\cal O}(1)$. This leads to a very
difficult signature with two hard jets and missing energy, which allows
discovery only for relatively low mirror quark masses \cite{mirror}.
For small values of
$R$, the mirror quarks become heavier than the SU(2) gauge bosons $W_H^{0,\pm}$,
but in this case their production cross section is also very small.

\begin{figure}[tb]
\centering
\epsfig{figure=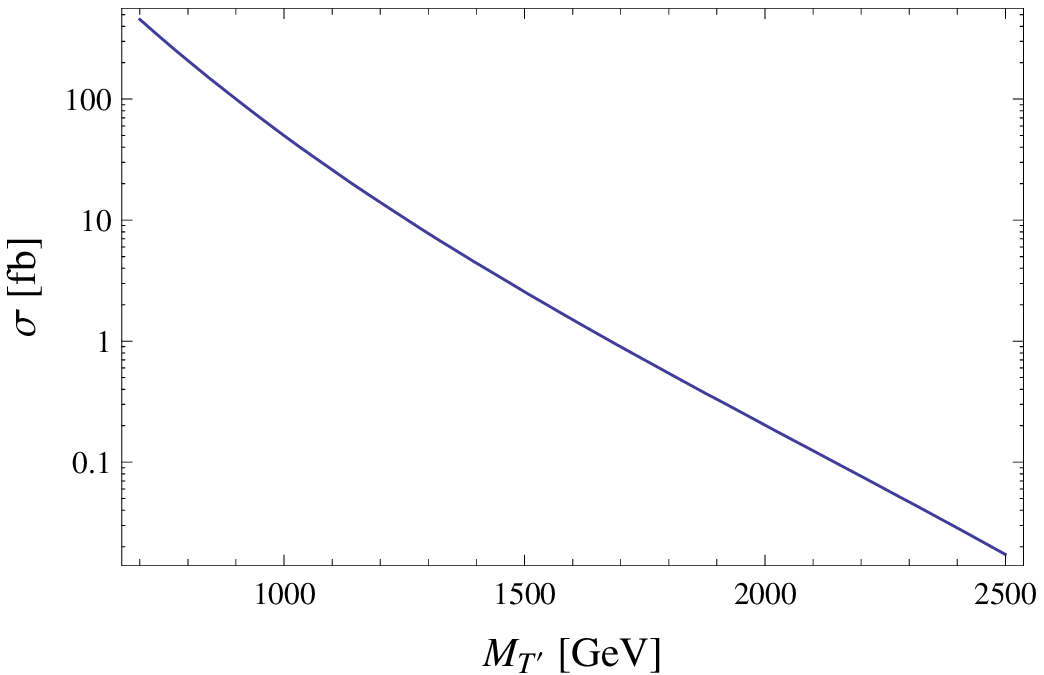, width=9.3cm}%
\vspace{-.3em}
\caption{LHC cross section for $T'\overline{T}{}'$
production, as a function of the $T'$ mass. 
The QCD and factorization scales have been
set to $M_{T^{\prime}}$, and the center-of-mass energy is
$\sqrt{s} =14$~TeV.}
\label{Tp}
\end{figure}

The most promising signal of an X-odd quark is expected from the additional top
quark partner $T'$, since it is predicted to be relatively light%
\footnote{Note that it is the $T'$ that cancels the quadratic divergences from
top quark loops to the Higgs mass parameter, and therefore its relatively low
mass is of central importance for the naturalness of the model.}.
As can be seen from Fig.~\ref{Tp}, the pair production cross section for $pp \to
T'\overline{T}'$ is sizable and can reach several 100 fb for ${\cal O}$(TeV)
masses of the $T'$. However, the decay signature $T'\overline{T}{}' \to
t\bar{t}\,B_H^0\,B_H^0$ is very similar to $t\bar{t}$ production in the SM,
so that a careful analysis is needed to disentangle signal from background
\cite{ttmiss}.

In contrast, very striking and clean signatures are expected from the production of
the scalar triplet $\phi_a$. 
In this model, $\phi_a^{0,\pm}$ is predicted to be relatively light, with a mass
that is about one order of magnitude below the scale $f$. Therefore the
production cross sections for the members of this triplet can be relatively
large. Since the triplet
$\phi_a^{0,\pm}$ is even under X-parity, but odd under T-parity, it decays through
the WZW term into SM gauge bosons:
\begin{align}
\phi_a^0 &\to  ZZ,\,\gamma\gamma,\gamma Z,\\
\phi_a^\pm &\to W^\pm Z,\, W^\pm \gamma. \label{wzw1}
\end{align}
Electroweak symmetry breaking can induce a small mass splitting between the
neutral and charged components of $\phi_a$. This mass splitting cannot be
calculated reliably since it can receive contributions from higher-dimension 
operators induced by the UV completion. Nevertheless, the leading contribution
to the mass splitting can be parametrically estimates to be of the order 
${\cal O}[g^4f^2/(4\pi^2)] \sim {\cal O}[g^4v^2]$. If we assume that
$\phi_a^\pm$ is heavier $\phi_a^0$ this opens up the additional decay channel
$\phi_a^\pm \to (W^\pm)^* \phi^0_a$  through a virtual $W$ boson.

\begin{figure}
\begin{center}
\epsfig{figure=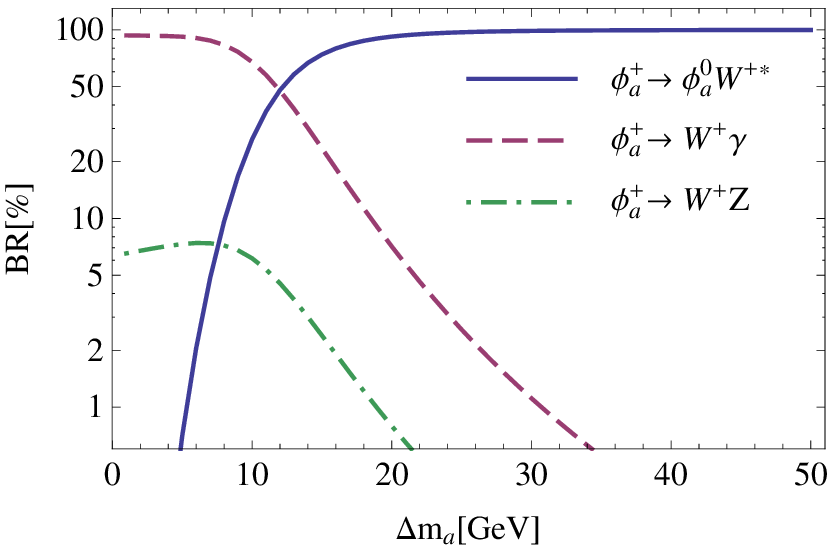, width=7.5cm}
% \begin{minipage}[b]{5cm}
% \psfig{figure=papa.ps, width=4.5cm}\\[1.5em]
% \psfig{figure=papa2.ps, width=4.5cm}\\[1em]
% \end{minipage}
\hspace{1em}
\epsfig{figure=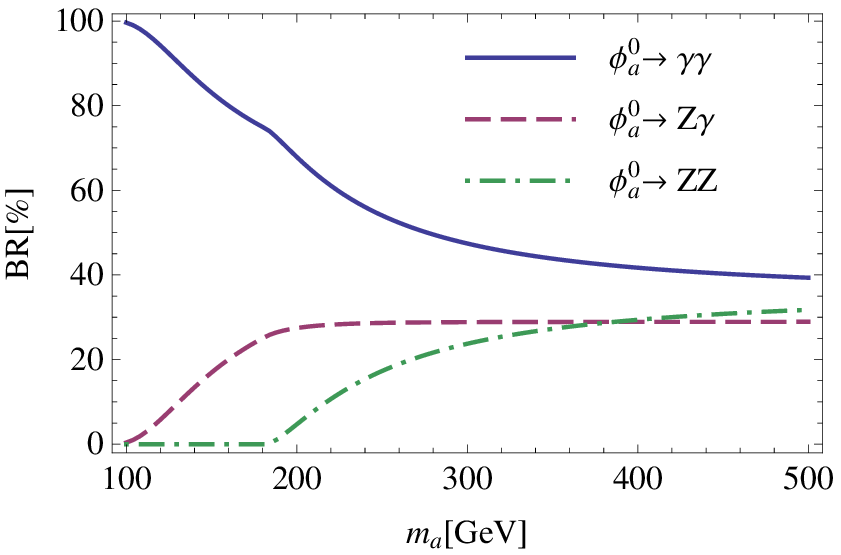, width=7.5cm}%
\end{center}
\vspace{-2em}
\caption{Left plot: Branching fractions of $\phi_a^+$ as a function of the mass splitting $\Delta m_a$. Right plot: Branching fractions of $\phi_a^0$ depending on the mass $m_a$.}
\label{apdecay}
\end{figure}

The left plot in figure \ref{apdecay} shows the branching fractions of the different decay modes
as a function of the mass splitting between $\phi_a^\pm$ and $\phi_a^0$. Once the
mass difference becomes larger than $15\gev$ the decays of $\phi_a^+$ through
a virtual $W^+$ will dominate over the WZW-term induced decays. 
The neutral $\phi_a^0$ dominantly decays into pairs of photons but also has sizeable
branching fractions into the $\gamma Z$ and $ZZ$ channels above the respective 
thresholds, as can be seen from the right plot in figure \ref{apdecay}. Note that the decay into $W^+W^-$ is absent because the WZW term
vanishes for the corresponding anomaly-free SU(2) subgroup.

\begin{figure}
\begin{center}
\begin{minipage}[b]{5cm}
\psfig{figure=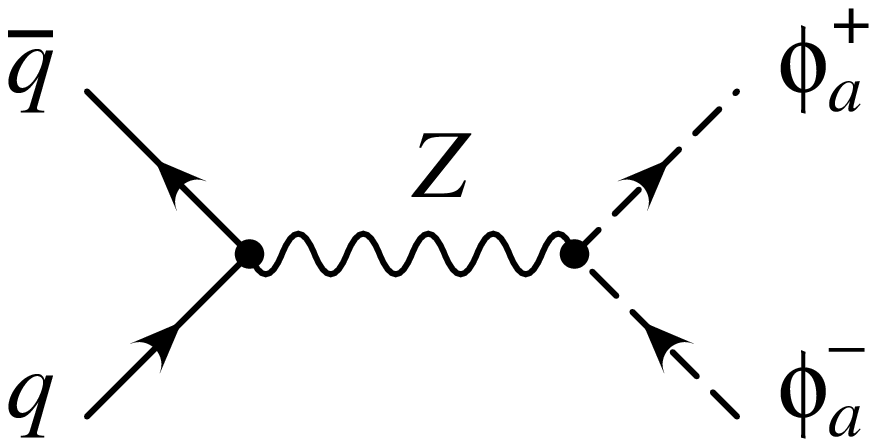, width=4.5cm}\\[1.5em]
\psfig{figure=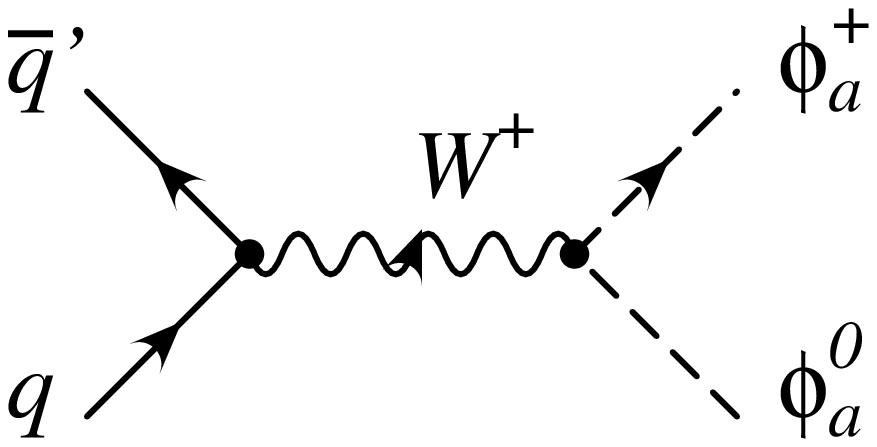, width=4.5cm}\\[1em]
\end{minipage}
\hspace{1em}
\epsfig{figure=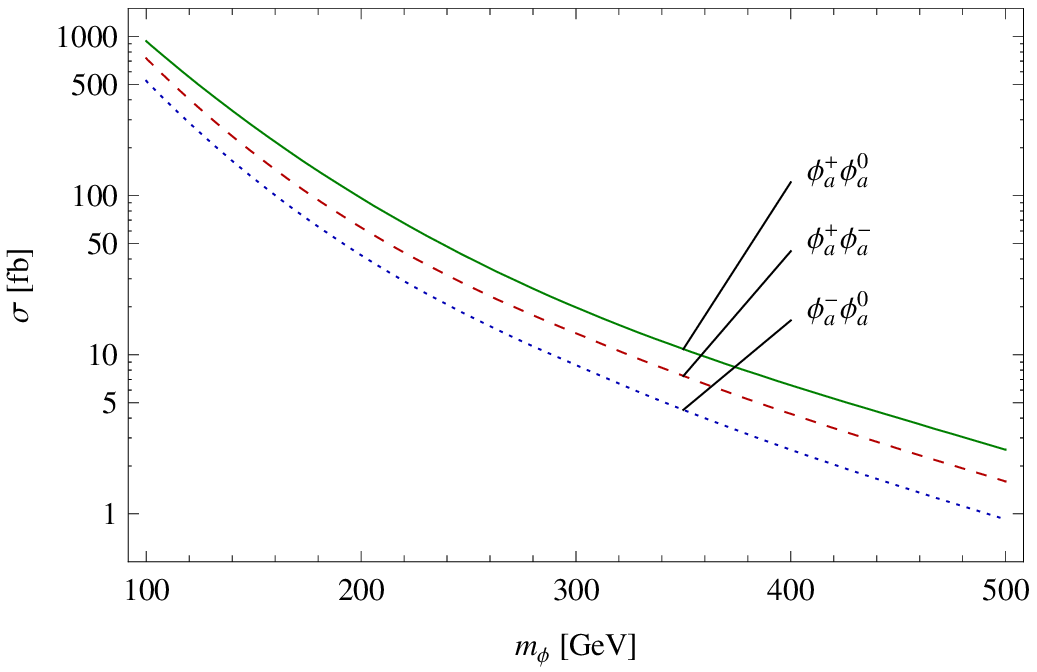, width=10cm}%
\end{center}
\vspace{-2em}
\caption{Pair production diagrams and
LHC cross sections for the particles in the lightest
scalar triplet, as a function of their mass. The factorization scale has been
chosen equal to $m_\phi$, and the center-of-mass energy is
$\sqrt{s} =14$~TeV.}
\label{phia}
\end{figure}

At the LHC, $\phi_a^{0,\pm}$ are mainly produced in pairs through
Drell-Yan-type processes with the Feynman diagrams shown in Fig.~\ref{phia}.
Since the WZW term is suppressed by several powers of $v/f$, single production
of $\phi_a^0$ or $\phi_a^\pm$ is completely negligible. Furthermore, production
of $\phi_a$ pairs through gluon fusion, which first occurs at one-loop level, is
also very small. Fig.~\ref{phia} shows the tree-level Drell-Yan production cross
sections, computed with the program  {\sc CompHEP 4.4} \cite{comphep},
using a model file generated with the help of  the {\sc LanHEP}
package \cite{lanhep}.

As evident from the figure, the largest cross section is obtained for mixed
pair production $pp \to \phi_a^0\,\phi_a^\pm$. 
Therefore striking
signals involving a $W$ boson and several photons are expected as a unique
signature of this process, with very little background from SM processes. 

In table \ref{phiatab} we show the signal rates for the most promising 
detection channels at the LHC involving charged leptons. We assume a mass
$m_a=300 \gev$, which is well beyond any direct detection bounds from the
Tevatron\footnote{The total production cross section for $\phi_a$ triplets at
the Tevatron is less then 1 fb in this case.}, and consider both the cases of small and large mass splitting between
$\phi_a^+$ and $\phi_a^0$.
\begin{table}
\begin{center}
% use packages: array
\begin{tabular}{|l|c|c||l|c|c|}\hline
$\Delta m_a = 5 \gev$ & $10 \tev$ & $14 \tev$ & $\Delta m_a = 20\gev$ & $10\tev$ & $14\tev$ \\ \hline \hline 
$l^+ \gamma \gamma \gamma \not{\!\!E} $ & 1.04 fb & 1.86 fb & $l^+ \gamma \gamma \gamma\gamma \not{\!\!E} $ & 0.47 fb & 0.84 fb \\ 
$l^+ l^+l^- \gamma \gamma \not{\!\!E} $ & 0.049 fb & 0.087 fb & $l^+ l^+l^- \gamma \gamma\gamma \not{\!\!E} $ & 0.038 fb & 0.068 fb \\ 
$l^+l^-\gamma\gamma \not{\!\!E}$ & 0.27 fb & 0.51 fb &  $l^+l^-\gamma\gamma\gamma\gamma \not{\!\!E}$& 0.053 fb & 0.10 fb\\ \hline
\end{tabular}
\caption{Signal rates for decays of $\phi_a^+\phi_a^0$ and $\phi_a^+ \phi_a^-$ pairs produced at LHC with $10 \tev$ and $14 \tev$ center of mass energy respectively. $l^\pm$ denotes either an electron or a muon. }
\label{phiatab}
\end{center}
\end{table}
We require at least one charged lepton in the final state from the leptonic
decay of the $W^+$. Events with three charged leptons are obtained from leptonic
decays of a $Z$ boson appearing either in the decays of $\phi_a^0$ or
$\phi_a^+$, while the events with two charged leptons come from decays of
$\phi_a^+ \phi_a^-$ pairs, where both $W$ bosons decay leptonically.  All
channels are very clean and essentially free of standard model backgrounds,
therefore these signals could be observed with a few fb$^{-1}$ at the LHC,
provided that instrumental backgrounds are under control. 

A large reduction of the signal rates occurs due to requiring leptonic decays of
the $W$ or $Z$ bosons. An alternative strategy could be to focus only on photons
to identify these signals. In table \ref{photontab} we present the signal rates
for events with more than two, three and four photons in the final state. 

\begin{table}
\begin{center}
% use packages: array
\begin{tabular}{|l|c|c||l|c|c|}\hline
$\Delta m_a = 5 \gev$ & $10 \tev$ & $14 \tev$ & $\Delta m_a = 20\gev$ & $10\tev$ & $14\tev$ \\ \hline \hline 
$ \gamma \gamma +X $ & 17.5 fb & 32.5 fb & $ \gamma \gamma +X $ & 15.1 fb & 28.2 fb \\ 
$ \gamma \gamma \gamma+X $ & 6.82 fb & 12.6 fb & $ \gamma \gamma \gamma+X $ & 9.31 fb & 17.4 fb \\ 
 &  &  &  $\gamma\gamma\gamma\gamma +X$& 4.20 fb & 7.87 fb \\ \hline
\end{tabular}
\caption{Multi-photon signals from decays of $\phi_a^+\phi_a^0$ and $\phi_a^+ \phi_a^-$ pairs produced at LHC with $10 \tev$ and $14 \tev$ center of mass energy respectively. }
\label{photontab}
\end{center}
\end{table}

The two-photon signal is most challenging to separate from the continuous
standard model di-photon background, as is well known from Higgs searches in
this channel. If possible at all, this would require an excellent mass
resolution to identify the $\phi_a^0 \rightarrow \gamma \gamma$ peak in the
invariant mass spectrum, and a rather high luminosity. Requiring a third photon
in the final state significantly reduces the standard model background. Here the
most important contributions are direct tri-photon production and events with
jets or electrons misidentified as photons. Assuming that the instrumental
backgrounds can be sufficiently suppressed, this channel has potential for an
early discovery at the LHC with only a few fb$^{-1}$.

\subsubsection*{6\hspace{1em} SUMMARY}

The implementation of an exact parity in a little Higgs model is an attractive
possibility to ensure good agreement with electroweak data and to yield a viable
dark matter candidate. Following an idea by Krohn and Yavin, a realistic 
model has been constructed that has an unbroken parity realized as an exchange
symmetry. It has been shown that this model is in agreement with electroweak
precision data and naturally accommodates electroweak symmetry breaking at a
scale of the order of 100 GeV.

In this contribution the collider phenomenology and possible discovery
of the model at the Large Hadron Collider (LHC) has been analyzed in detail.
It was found that many particles with reach of the LHC are predicted, but
discovery is challenging for many channels due to low cross sections and
signatures with large backgrounds. The most promising signal stems from the
production of light parity-even SU(2) triplet scalars, which have masses near
the electroweak scale. These triplet scalars decay into pairs of Standard Model
gauge bosons, leading to striking signatures involving multiple photons and 
leptons from $W$ and $Z$ bosons.

A few open question remain for further study. While the heavy U(1) gauge boson
$B_H^0$ is a promising dark matter candidate with the correct quantum numbers,
it remains to be checked whether the observed relic dark matter density could be
obtained within this model.
Furthermore, for successful electroweak symmetry breaking, the Higgs potential
must contain a complex parameter, which might lead to potentially
observable CP violating effects. The CP violating parameter could also play a
role for electroweak baryogenesis.

\end{document}